\documentclass[acus]{pac99}
\usepackage{epsfig}
\newcommand{\bit}{\begin{Itemize}}
\newcommand{\eit}{\end{Itemize}}
\begin{document}

\thispagestyle{empty}

\onecolumn

\begin{flushright}
{\large
SLAC--PUB--8233\\
August 1999\\}
\end{flushright}

\vspace{.8cm}

\begin{center}

{\LARGE\bf
Effect of Insertion Devices in SPEAR~3\footnote
{\normalsize{Work supported by
Department of Energy contract  DE--AC03--76SF00515.}}}

\vspace{1cm}

\large{
J.~Corbett and
Y.~Nosochkov \\
Stanford Linear Accelerator Center, Stanford University,
Stanford, CA  94309\\
}

\end{center}

\vfill

\begin{center}
{\LARGE\bf
Abstract }
\end{center}

\begin{quote}
\large{
The SPEAR~3 upgrade lattice will provide much reduced
beam emittance to increase the brightness of synchrotron
radiation beams from wigglers and undulators. Seven
existing insertion devices will be used in the lattice.
In this paper we review the wiggler parameters,
outline the wiggler compensation scheme, and evaluate
wiggler effect on the optics and dynamic aperture.
}
\end{quote}

\vfill

\begin{center}
\large{
{\it Presented at the 1999 IEEE Particle Accelerator Conference (PAC99)\\
New York City, New York, March 29 -- April 2, 1999} \\
}
\end{center}

\newpage

\pagenumbering{arabic}
\pagestyle{plain}

\twocolumn

\title{
EFFECT OF INSERTION DEVICES IN SPEAR~3\thanks
{Work supported by the Department of Energy
Contract DE--AC03--76SF00515.}
}

\author{
\underbar{J.~Corbett\thanks{E-mail: corbett@slac.stanford.edu.}} 
and Y.~Nosochkov \\
Stanford Linear Accelerator Center, Stanford University,
Stanford, CA 94309
}
\maketitle

\begin{abstract}
The SPEAR~3 upgrade lattice will provide much reduced
beam emittance to increase the brightness of synchrotron
radiation beams from wigglers and undulators. Seven 
existing insertion devices will be used in the lattice.
In this paper we review the wiggler parameters, 
outline the wiggler compensation scheme, and evaluate
wiggler effect on the optics and dynamic aperture.
\end{abstract}

\vspace{-1mm}
\section{INTRODUCTION}

Many of the drift sections in SPEAR~3 will contain wigglers or undulators. 
In the present machine, there are 7 insertion devices (ID). In this 
paper, we review the magnetic parameters for each of these ID's, 
estimate linear and higher order wiggler effects, evaluate radiation 
effects, and discuss correction of wiggler focusing using quadrupole 
trim windings. Tracking results show that dynamic aperture reduction due 
to wiggler effects is modest. All optics calculations were done with the 
MAD code \cite{mad}, and the tracking simulations with the LEGO 
code \cite{lego}.
\vspace{-1mm}
\section{WIGGLER PARAMETERS, EFFECTS AND MODELING}
In SPEAR~3, the horizontally deflecting ID's reside in dispersion free
straights where the unperturbed $\beta$ functions are
$\beta_{x}$/$\beta_{y}=$10.1/4.8 m. The magnetic field parameters for 
each ID are given in Table~1, where $N$ is the number of periods, 
$\lambda$ the period length, $B_o$ the peak field.\footnote{
Beamline 5 can use five separate devices:
(1) $N$=10, $\lambda$=0.18,
(2) $N$=15, $\lambda$=0.12, (3) $N$=24, $\lambda$=0.07,
(4) $N$=30, $\lambda$=0.06, and
(5) elliptically polarized undulator (EPU).
Here, we chose the $N$=10, $\lambda$=0.18 wiggler.}
Future devices are likely to be $\sim$10-period wigglers
similar to Beamlines 9 and 11. These devices will replace
the wigglers on Beamlines 4 and 7 or illuminate new
beam lines.

The main optical effects of the wigglers on the SPEAR 3
lattice are also summarized in Table~1. The parameter definitions and 
relevant formulas are listed below:
\vspace{-1mm}
\begin{tabbing}
TTTTTTTTTTTTTTTTT \= TTTTTTTTTTTTT \= \kill
Bend radius \> $\rho_{o}(m)=3.3356 \frac{E(GeV)}{B_{o}(T)}$ \\
Wiggler parameter \> $K = 93.44 B_{o}(T) \lambda(m)$ \\
Integrated focusing~\cite{wiedmn} \> $\int{k_{y}ds} =
\frac{N \lambda} {2\rho_{o}^{2}}$ \\
Linear tune shift \> $\Delta Q_y = \frac{\beta_{y}}{4\pi}
\int{k_{y}ds}$ \\
Maximum $\beta$-beat \> $\frac{\Delta \beta_{y}}{\beta_{y}} =
\frac{\beta_{y}\int{k_{y}ds}}{2sin2\pi Q_{y}}$ \\
Amplitude dependent \> $\Delta Q^{oct}_{y}=
\epsilon_{y}\frac{\pi N \beta_{y}^{2}}{4\lambda \rho_{o}^{2}}
[1+\frac{2}{3}(\frac{N\lambda}{2\beta_{y}})^{2}$ \\
tune shift~\cite{smith} \> \>
$+\frac{1}{5}(\frac{N\lambda}{2\beta_{y}})^{4}]$
\end{tabbing}
where $Q_{y}$=5.23 is the vertical betatron tune, $\epsilon_{y}$ the 
vertical emittance, and $E$=3 GeV the beam energy.

\begin{table}[tb]
\small
\begin{center}
\caption{Wiggler parameters and main optics effects.}
\vspace{-2mm}
\medskip
\begin{tabular}{lccccccc}
\hline
\textbf{Beamline} & \textbf{4,7} & \textbf{5}  & 
\textbf{6} & \textbf{9} & \textbf{10} & \textbf{11} \\
\hline
$N$ & 4 & 10 & 27 & 8 & 15 & 13 \\
$\lambda$ (cm) & 45 & 18 & 7 &
26 & 13 & 17.4 \\
$B_o$ (T) & 1.8 & 0.9 & 1.3 & 1.93 &
1.45 & 1.98 \\
$\rho_{o}$ (m) & 5.56 & 11.12 & 7.70 &
5.18 & 6.90 & 5.05 \\
$K$ & 75.7 & 15.1 & 8.5 & 46.9 & 17.6 & 32.2 \\
$\int k_{y}ds$ ($m^{-1}$) & .029 & .007 &
.016 & .039 & .021 & .044 \\
$\Delta Q_y$ & .011 & .003 & .006 & .015 & .008 & .017 \\
$\Delta \beta_{y}/\beta_{y}$ (\%) & 7.0 & 1.8 &
3.9 & 9.4 & 5.0 & 10.7 \\
$\Delta Q^{oct}_{y}$[$10^{-4}$] &
0.4 & 0.6 & 9.1 & 1.6 & 3.4 & 4.1 \\
\hline
\end{tabular}
\label{tab:param}
\end{center}
\vspace{-7mm}
\end{table}

For a flat horizontal wiggler the field components can be expressed 
as \cite{halbach}:
\begin{eqnarray}
B_{y}=B_{o}\cosh (ky)\cos (kz),
\label{eq:By} \\
B_{z}=-B_{o}\sinh (ky)\sin (kz),
\label{eq:Bz}
\end{eqnarray}
where $k=2\pi/\lambda$, $y$ and $z$ are the vertical and 
longitudinal coordinates, and $z$=0 the center of a pole. 
The integral of the $B_{y}$ field 
(Eqn.~\ref{eq:By}) with respect to the reference
trajectory vanishes over each wiggler period. This leads to self
compensation of the optics effects. For instance, the horizontal 
orbit and dispersion generated by $B_{y}$ are fully localized within 
each period. The amplitude of the orbit and dispersion
is proportional to $\lambda^{2}B_{o}$, and the increase
in beam path length is proportional to $\lambda^{3}B_{o}^{2}N$. 
Hence, the largest orbit oscillations are produced in the 
wigglers with the longest poles. The total increase of the path 
length for all 7 wigglers is 191 $\mu$m.

Taking into account the horizontal orbit oscillations of the beam,
the $B_{z}$ field (Eqn.~\ref{eq:Bz}) can be decomposed into 
longitudinal and horizontal components on the beam orbit. Although 
the integral of the longitudinal component vanishes over each period, 
one can can find a non-zero integral of the field horizontal to
the beam trajectory \cite{wiedmn}:
\begin{equation}
\int_{o}^{N\lambda} B_{\tilde{x}}dz=-\frac{N\lambda B_{o}^{2}}
{2B\rho}(y+ \frac{2}{3}k^{2}y^{3}+\frac{2}{15}k^{4}y^{5}+\ldots),
\label{eq:Bx}
\end{equation}
where label $\tilde{x}$ refers to the local horizontal axis
with respect to the oscillating beam orbit. 

The first term in Eqn.~\ref{eq:Bx} ($\propto{y}$) is similar to the 
vertical gradient in quadrupoles and thus results in vertical focusing, 
namely the vertical tune shift $\Delta Q_y$ and vertical betatron beat 
$\Delta \beta_{y}/\beta_{y}$. The total linear tune shift is 
$\Delta Q_y$=0.071 for all 7 wigglers. The pattern of the $\beta$-beat 
depends on the distribution of wigglers in the ring. The $\sim$$90^{o}$ 
vertical phase advance per cell helps to minimize beta perturbations 
from a pair of wigglers if they are located on either side of the same cell. 
All 7 wigglers generate about $\pm$25\% vertical $\beta$-beat in the 
ring without correction.

The first non-linear wiggler effect comes from the second term in 
Eqn.~\ref{eq:Bx} ($\propto{y^{3}}$). This octupole-like field generates
quadratically increasing vertical tune shift with $y$-amplitude.
For devices of similar length ($N\lambda$) the effect is proportional 
to $1/\lambda^2$. In Table~1, this tune shift was evaluated for the 
maximum vertical emittance ($\epsilon_{y}$=7.5 mm$\cdot$mrad) defined 
by the smallest vacuum chambers at Beamlines 6 and 11 
($y$=6 mm, $\beta_{y}$=4.8 m). The maximum total tune shift for 7 
wigglers is $\Delta Q^{oct}_{y}$=0.002. This value is significantly 
smaller than the amplitude dependent tune shift caused by the 
sextupoles and was considered acceptable. Tracking simulations
with wiggler effects confirm that the vertical dynamic aperture
stays well outside the physical wiggler aperture.

For optics calculations, each wiggler pole was simulated by 
a shorter pole with constant vertical field and with gaps between the 
poles \cite{wiedmn}. In this model, the pole field, $B_{p}=B_{o} \pi/4$, 
and pole length, $L_{p}=4\lambda /\pi^2$, were set to produce the same 
bending and focusing effect as the actual wiggler field. Therefore, 
the model yields the correct linear tune shift, $\beta$-beat, and the 
net orbit and dispersion. For tracking studies, higher order multipole 
fields were added to simulate the effects of non-linear wiggler fields 
on dynamic aperture.
\vspace{-1mm}
\section{RADIATION EFFECTS}
Synchrotron radiation and quantum excitation in wigglers increase 
the beam energy loss and change the equilibrium emittance and the beam 
energy spread. For a single ID, the relative increase in energy loss 
per turn is \cite{ropert}:
\begin{equation}
\frac{U-U_{o}}{U_{o}}=\frac{N\lambda\rho_{b}}{4\pi\rho_{o}^2},
\label{eq:dU}
\end{equation}
where $\rho_{b}$=7.86 m is the bend radius in the main dipoles, 
and $U_{o}$=913 keV/turn (without ID's). The energy loss with all 7
wigglers at full field increases by 23\% to 1124 keV/turn. The total 
radiated wiggler power is 225 kW from a 200 mA beam.

The relative change in the beam energy spread due to each ID is given 
by \cite{ropert}:
\begin{equation}
(\frac{\sigma_E}{\sigma_{E}^{o}})^{2}=
[1 + \frac{2N\lambda\rho_{b}^{2}}{3\pi^{2}\rho_{o}^{3}}] / 
[1 + \frac{N\lambda\rho_{b}}{4\pi\rho_{o}^{2}}].
\label{eq:sigE}
\end{equation}

\begin{table}[b]
\vspace{-5mm}
\small
\begin{center}
\caption{Change in energy loss and energy spread due to wigglers.}
\medskip
\begin{tabular}{lcccccc}
\hline
\textbf{Beamline} & \textbf{4,7} & \textbf{5} & \textbf{6} & 
\textbf{9} & \textbf{10} & \textbf{11} \\
\hline
$\frac{U-U_{o}}{U_{o}} (\%)$ & 3.64 & 0.91 & 1.99 & 4.84 & 2.56 & 5.54 \\
\hline
$\frac{\sigma_{E}-\sigma_{E}^{o}}{\sigma_{E}^{o}} (\%)$ & 
0.35 & -0.18 & -0.13 & 0.66 & -0.04 & 0.84 \\
\hline
\end{tabular}
\label{tab:rad}
\end{center}
\vspace{-5mm}
\end{table}

With 7 wigglers at full field, the net energy spread on the beam increases
by only 1.6$\%$, from $\sigma_{E}^{o}/E$=0.097\% to 0.0986\%.
The effects of individual ID's are listed in Table~2. Emittance calculations
show that 7 wigglers reduce the horizontal emittance from 18.2 nm$\cdot$rad
to 15.3 nm$\cdot$rad. 
\vspace{-1mm}
\section{OPTICS COMPENSATION}
The wiggler focusing from 7 wigglers at full field generates 
$\pm$25\% vertical $\beta$-beat and a linear vertical tune shift of 0.071.
This distortion breaks the periodicity of the $\beta$-functions and 
phase advance between sextupoles, which could lead to stronger sextupole
resonances and reduced dynamic aperture. 

\begin{figure}[t]
\includegraphics{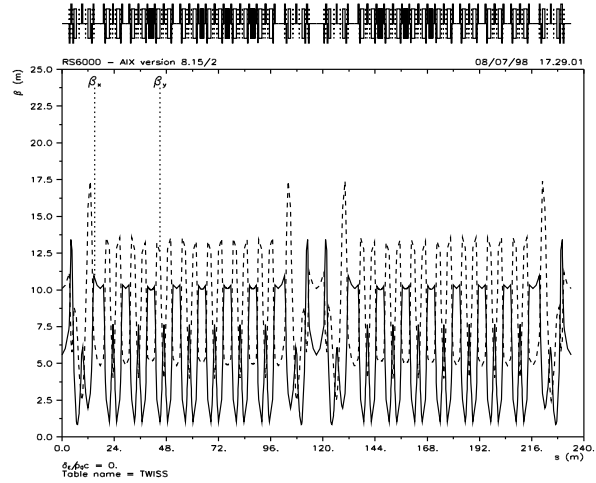}
\vspace{65mm}
\caption{SPEAR~3 $\beta$ functions with 7 wigglers (black boxes on the top) 
after correction.}
\label{fig:beta}
\vspace{-4mm}
\end{figure}

Several options are possible to correct the wiggler focusing.
The preferred scheme is to locally compensate the distortion of
$\beta$-functions and phase advance with four QF and four QD independent 
quadrupole trims located to either side of each ID. With this correction, 
the phase advance stays identical in all cells, the global tune adjustment
is not necessary and the distortion of lattice functions at sextupoles is
minimal. The required independent quadrupole trims are less than 10\%. 
This correction scheme has been designed into the SPEAR~3 lattice and was 
used in tracking studies described in the following section.
The SPEAR~3 $\beta$ functions after correction of 7 wigglers are shown
in Fig.~\ref{fig:beta}. 

\vspace{-1mm}
\section{EFFECT ON DYNAMIC APERTURE}
Effects of IDs on dynamic aperture arise from reduced symmetry and lattice
periodicity, distortion of $\beta$-functions and betatron phase, non-linear 
wiggler fields, wiggler field errors and misalignment. In tracking 
simulations, we studied the following effects of wigglers on dynamic 
aperture:

\vspace{1mm}
\noindent
- Linear wiggler focusing, \\
\noindent
- Effect of systematic multipole field errors in wigglers, \\
\noindent
- Effect of intrinsic non-linear wiggler field. \\
\vspace{0mm}
 
\begin{table}[tb]
\begin{center}
\caption{Multipole fit to the measured systematic multipole field error in the
wiggler on Beamline 11 ($G/cm^{n-2}$).}
\medskip
\begin{tabular}{lcccccc}
\hline
\boldmath{$n$} & \textbf{1} & \textbf{2} & \textbf{3} & \textbf{4} & 
\textbf{5} & \textbf{6} \\
\hline
$b_{n}L$ & 20.3 & -6.49 & -15.5 & -37.2 & 5.96 & 33.2 \\
$a_{n}L$ & 19.4 &  17.9 & -35.4 &  26.4 & 3.64 & -65.0 \\
\hline
\end{tabular}
\label{tab:mult1}
\end{center}
\vspace{-7mm}
\end{table}
\begin{table}[ht]
\begin{center}
\medskip
\begin{tabular}{lcccccc}
\hline
\boldmath{$n$} & \textbf{7} & \textbf{8} & \textbf{9} & \textbf{10} & 
\textbf{11} & \textbf{12} \\
\hline
$b_{n}L$ & 2.73 & -7.40 & -1.05 & 0.457 & 0.079 & 0.001 \\
$a_{n}L$ & 13.9 & 33.3 & -3.55 & -5.57 & 0.227 & 0.297 \\
\hline
\end{tabular}
\label{tab:mult2}
\end{center}
\vspace{-7mm}
\end{table}

Without correction of the wiggler focusing, the vertical dynamic aperture
reduces by about 20\% with all wigglers active. With corrected wiggler
focusing, linear wiggler effects alone do not reduce the aperture.

The study of wiggler multipole errors was based on measured systematic
field errors in the wiggler on Beamline 11. A numerical multipole fit to
the measured data was performed and the results for the integral of
normal and skew systematic multipole field are shown in Table~3, where
`$n$' is the multipole order starting with dipole, $L=N\lambda$, and
$a_{n}$, $b_{n}$ are the skew and normal field coefficients defined as
\begin{equation}
B_{y}+iB_{x}=\sum_{n}(b_{n}+ia_{n})(x+iy)^{n-1}.
\label{eq:BxBy}
\end{equation}

\begin{figure}[tb]
\includegraphics{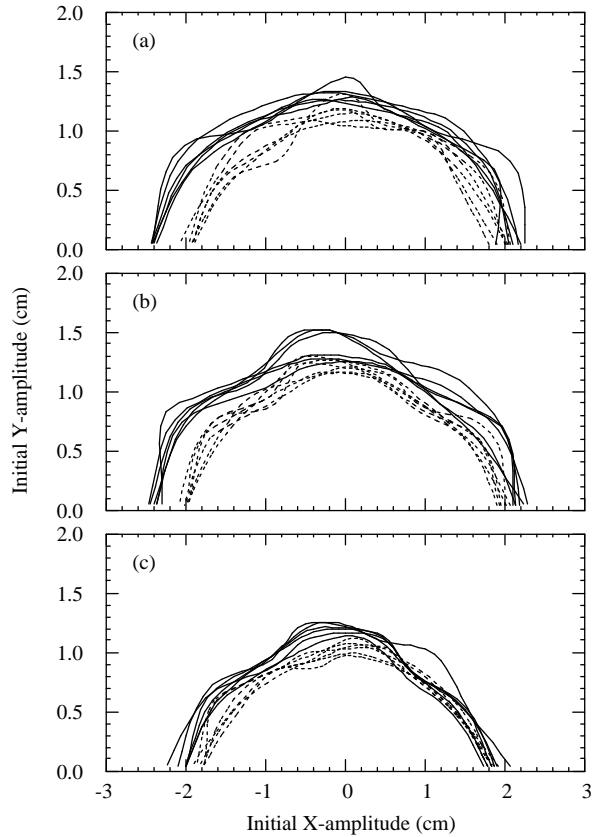}
\vspace{115mm}
\caption{Dynamic aperture for $\delta$=0 (solid) and 3\% (dash)
off-momentum oscillations and 6 seeds of random machine errors:
(a) no wigglers, (b) 7 wigglers with corrected focusing but w/o
errors, (c) 7 wigglers with corrected focusing and with systematic
multipole errors.}
\label{fig:aper}
\vspace{-2mm}
\end{figure}

The effect of 7 wigglers on dynamic aperture is shown in
Fig.~\ref{fig:aper}, with the aperture calculated at the ID locations
($\beta_{x}$/$\beta_{y}=$10.1/4.8 m) for machine simulations with all
errors in the ring magnets. Without multipole errors in the wigglers but
with corrected wiggler focusing, the aperture is not affected.
With the systematic multipole errors included in the 7 wigglers,
the dynamic aperture reduction is about 10\% (2 mm).

The wiggler-intrinsic high order terms (Eqn.~\ref{eq:Bx}) were not
included in the previous simulation, but can further 
reduce the dynamic aperture. This field does not have the usual form 
of ($x$-$y$) multipole expansion normally used in lattice codes. 
To accurately study the effect of the octupole-like and higher order 
terms in Eqn.~\ref{eq:Bx}, one needs a symplectic model of this
field. We used an approximation to evaluate the main effect 
of these terms. Since the non-linear $B_{\tilde{x}}$ field generates 
kicks only in $y$-plane, the effect on dynamic aperture is expected to be 
mostly in the vertical plane, as it was observed in simulations for
other machines \cite{ropert}. 

One can notice that the $B_{\tilde{x}}$ field can be reproduced at $x$=0
by the normal field of Eqn.~\ref{eq:BxBy} ($a_n$=0). Using this approach, 
we used normal systematic multipoles in the form of 
$B_{y}+iB_{x}=\sum b_{n}(x+iy)^{n-1}$ to produce the non-linear terms of
the wiggler field $B_{\tilde{x}}$ on the vertical axis (Eqn.~\ref{eq:Bx}), 
and performed tracking to evaluate vertical aperture near $x$=0.
Simulation of the octupole-like and dodecapole-like ($\propto{y^{5}}$) 
fields with 7 wigglers showed only modest reduction of vertical 
dynamic aperture, from 11 mm to 8-9 mm. This aperture still remains well 
outside the 6 mm wiggler physical aperture. 
\vspace{-1mm}
\section{SUMMARY}
Tracking simulations with 7 wigglers in SPEAR~3 predict small impact
on dynamic aperture when linear optics effects are locally compensated
with individually controlled quadrupole magnets. The 18 mm horizontal 
dynamic aperture is sufficient for 10 mm injection oscillations and long 
Touschek lifetime, and the vertical aperture remains well outside the
6 mm wiggler physical aperture. With all wigglers active,
the increase in energy spread is negligible ($<$2\%), and the beam
horizontal emittance decreases to about 15.3 nm$\cdot$rad at 3 GeV.

\vspace{-1mm}
\section{ACKNOWLEDGEMENTS}
We would like to thank M.~Cornacchia, H.~Wiedemann and 
B.~Hettel for useful discussions, suggestions and guidance.

\end{document}